\documentclass[12pt]{article}
\pdfoutput=1
\usepackage{draft,tikz-cd,shuffle,float}
\usetikzlibrary{decorations.pathmorphing}	
\usetikzlibrary{decorations.markings}
\tikzset{
	photon/.style={decorate, decoration={snake}, draw=red},
	electron/.style={draw=blue, postaction={decorate},
		decoration={markings,mark=at position .55 with {\arrow[draw=blue]{>}}}},
	gluon/.style={decorate, draw=blue,
		decoration={coil,amplitude=4pt, segment length=4pt}} ,
	vector/.style={decorate, decoration={snake}, draw},
	provector/.style={decorate, decoration={snake,amplitude=2.5pt}, draw},
	antivector/.style={decorate, decoration={snake,amplitude=-2.5pt}, draw},
	fermion/.style={draw=black, postaction={decorate},
		decoration={markings,mark=at position .55 with {\arrow[draw=black]{>}}}},
	fermionbar/.style={draw=black, postaction={decorate},
		decoration={markings,mark=at position .55 with {\arrow[draw=black]{<}}}},
	fermionnoarrow/.style={draw=black},
	scalar/.style={dashed,draw=black, postaction={decorate},
		decoration={markings,mark=at position .55 with {\arrow[draw=black]{>}}}},
	scalarbar/.style={dashed,draw=black, postaction={decorate},
		decoration={markings,mark=at position .55 with {\arrow[draw=black]{<}}}},
	scalarnoarrow/.style={dashed,draw=black},
	electron/.style={draw=black, postaction={decorate},
		decoration={markings,mark=at position .55 with {\arrow[draw=black]{>}}}},
	bigvector/.style={decorate, decoration={snake,amplitude=4pt}, draw},
	background/.style={dashed,draw=black, postaction={decorate},
		decoration={markings,mark=at position 1 with {\arrow[draw=black]{<>}}}},
}

\tikzstyle{block} = [draw, rectangle, 
minimum height=3em, minimum width=6em]

\def\ma{\mathcal}
\def\mb{\mathbf}
\begin{document}

\begin{titlepage}

\begin{center}

\hfill \\
\hfill \\
\vskip 1cm

\title{Celestial Berends-Giele current}

\bigskip

\author{Yi-Xiao Tao$^{a}$}

\bigskip

\address{${}^a$Department of Mathematical Sciences, Tsinghua University, Beijing 100084, China}

\email{taoyx21@mails.tsinghua.edu.cn}

\end{center}

\vfill

\begin{abstract}
Celestial amplitude plays an important role in the understanding of holography. Computing celestial amplitudes by recursion can deepen our understanding of the structure of celestial amplitudes. As an important recursion method, the Berends-Giele (BG) currents on the celestial sphere are worth studying. In this paper, we study the celestial BG recursion and utilize this to calculate some typical examples. We also explore the OPE behavior of celestial BG currents. Moreover, we generalize the ``sewing procedure" for BG currents to the celestial case.

\end{abstract}

\vfill

\end{titlepage}

\tableofcontents

\newpage
\section{Introduction}
Recursion methods are very important in amplitudes. We can obtain higher point amplitudes from the lower point ones using recursion relations, which means that we can convert the complication calculation of higher point amplitudes to the much easier calculation of lower point ones. In the past several decades, the most popular recursion relation is the Britto-Cachazo-Feng-Witten (BCFW) recursion \cite{Britto:2004ap,Britto:2005fq}, which is based on the complex-shift method and the residue theorem. Another on-shell recursion, which is raised even before the complex-shift method, is the Cachazo-Svrcek-Witten (CSW) expansion \cite{Cachazo:2004kj,Risager:2005vk,Elvang:2008vz}. For a more comprehensive introduction, see \cite{Elvang:2013cua}. Apart from the on-shell recursion, there is also an off-shell, or a semi-on-shell, recursion relation based on the Berend-Giele (BG) currents \cite{Berends:1987me,Berends:1988zp,Berends:1988zn}. The definition of the BG currents is amplitudes with one leg off-shell. And the BG recursion can be obtained by evaluating the equation of motion. In other words, the origin of the BG recursion can be regarded as Feynman rules, which means this recursion is very convenient to be generalized to other cases. In recent years, there have been some new developments in the study of BG currents \cite{Armstrong:2022mfr,Chen:2023bji,Frost:2020eoa,Mafra:2015vca,Mafra:2016ltu,Mafra:2016mcc,Tao:2022nqc,Wu:2021exa}.

Inspired by the holographic duality of AdS spacetime, we wonder if we can regard the amplitudes as CFT correlators. In fact, after we change the basis of amplitudes, we will obtain CFT correlators on the ``celestial sphere" \cite{Pasterski:2017kqt}. A superamplitude version can be found in \cite{Brandhuber:2021nez}. There are also some results about celestial twistor amplitudes \cite{Brown:2022miw} and celestial string amplitudes \cite{Donnay:2023kvm,Stieberger:2018edy}. A natural question is, can we find some recursion relations for the celestial amplitudes? In \cite{Pasterski:2017ylz}, the authors show an example of BCFW for 4-pt MHV gluon celestial amplitudes. For n-pt MHV gluon celestial amplitudes, a formula based on BCFW is given in \cite{Guevara:2019ypd,Hu:2022bpa}. In appendix \ref{A}, we show a 3D BCFW method for celestial amplitudes as a generalization. 

However, the off-shell method on the celestial sphere still needs to be studied. In \cite{Pasterski:2020pdk}, the authors show an off-shell prescription for scalars. Furthermore, in \cite{Melton:2021kkz}, the author discusses the Feynman rules for celestial scalar amplitudes. In this paper, we will show how to construct BG currents for celestial amplitudes utilizing the gluon case as an example.

This paper will be organized as follows. In section \ref{sec2}, we will review some basic concepts including celestial amplitudes and BG currents. In section \ref{sec3}, we derive the celestial BG currents for gluon amplitudes and show how to do the recursion. Moreover, we reproduce the result of 3-pt celestial MHV gluon amplitude in the collinear region (in the Klein space with signature ($-+-+$) )\cite{Pasterski:2017ylz} and the soft region (in the Minkowski space with signature ($-+++$)) \cite{Chang:2022seh}. Finally, we generalize the ``sewing procedure" of BG currents to celestial BG currents in section \ref{sec4}.

\section{Preliminaries}\label{sec2}
In this section, we will review some basic concepts. For a more comprehensive review of the celestial amplitudes, see \cite{Raclariu:2021zjz}.
\subsection{Celestial amplitude}
Celestial amplitudes can be obtained by expanding position space amplitudes $M(X_{i})$ with respect to the conformal primary wave functions $\phi_{\Delta,s}^{\pm}(z,\bar{z};X)$ (or shadow conformal primary wave functions $\tilde{\phi}^{\pm}_{\Delta,s}(z,\bar{z};X)$ instead of plane waves:
\ie
\ma{M}^{\Delta_{i}}_{s_{i}}(z_{i},\bar{z}_{i})=(\prod_{j=1}^{n} \int d^4X_{j})( \prod_{j=1}^{k}\phi^{+}_{\Delta_{j},s_{j}}(z_{j},\bar{z}_{j};X_{j}))(\prod_{j=k+1}^{n} \tilde{\phi}^{-}_{\Delta_{j},s_{j}}(z_{j},\bar{z}_{j};X_{j}))M(X_{i}).
\fe
Here $\Delta$ and $s$ are the conformal dimension and the spin (or helicity) respectively, and $\pm$ labels whether the particle is outgoing or incoming. 
The coordinates $(z_{i},\bar{z}_{i})$ on the celestial sphere can be connected to massless on-shell momenta $q_{i}^{\mu}$ through
\ie
q^{\mu}=\omega\hat{q}^{\mu}=\omega(1+z\bar{z},z+\bar{z},-i(z-\bar{z}),1-z\bar{z})
\fe
and for massive on-shell momentum $p^{\mu}$ we have
\ie
p^{\mu}=m\hat{p}^{\mu}=\frac{m}{2y}(1+y^2+w\bar{w},w+\bar{w},-i(w-\bar{w}),1-y^2-w\bar{w})
\fe
Note that $\omega$ and $m$ are always positive, which means that whether the particle is incoming or outgoing is very important in this context. Here, we choose to expand the wave functions of the incoming particles in the conformal primary basis while the outgoing particles in the shadow conformal primary basis \cite{Chang:2022seh}. Now we review some wave functions commonly used. 

For scalars, the conformal primary wave functions are given by
\ie
\phi_{\Delta}^{\pm}(z,\bar{z};X)=\int_{0}^{\infty}d\omega\ \omega^{\Delta-1}e^{\pm i\omega\hat{q}\cdot X-\epsilon\omega}=\frac{(\mp i)^{\Delta}\Gamma(\Delta)}{(-\hat{q}\cdot X\mp i\epsilon)^{\Delta}}\propto \frac{1}{(-\hat{q}\cdot X\mp i\epsilon)^{\Delta}}
\fe
for massless scalars and
\ie
\phi_{\Delta,m}^{\pm}(z,\bar{z};X)=\int\frac{d^3 \hat{p}'}{\hat{p}^{\prime0}}\frac{1}{(-\hat{q}\cdot\hat{p}')^{\Delta}}e^{\pm im\hat{p}'\cdot X}=\int\frac{dy}{y^3}dwd\bar{w}(\frac{y}{y^2+|z-w|^2})^{\Delta}e^{\pm im\hat{p}(y,w,\bar{w})\cdot X}
\fe
for scalars with mass $m$. For massive scalars, shadow conformal primary wave functions can be treated as conformal primary wave functions \cite{Pasterski:2017kqt}; while for massless scalars, we have
\ie
\tilde{\phi}^{\pm}_{\Delta}(z,\bar{z};X)\propto(-X^2)^{\Delta-1}\phi^{\pm}_{\Delta}(z,\bar{z};X)
\fe
For spin-1 particles, things are more complicated since we need to include the gauge terms. For massless spin-1 particles, the conformal primary wave function is
\ie
A_{\mu s}^{\Delta,\pm}(\hat{q};X)=&\frac{\Delta-1}{\Delta}\frac{\partial_{s}\hat{q}_{\mu}}{(-\hat{q}\cdot X\mp i\epsilon)^{\Delta}}+\partial_{\mu}\frac{\partial_{s}\hat{q}\cdot X}{\Delta(-\hat{q}\cdot X\pm i\epsilon)^{\Delta}}\\
=&V_{\mu s}^{\Delta,\pm}(\hat{q};X)+\partial_{\mu}\alpha_{s}^{\Delta,\pm}(\hat{q};X),
\fe
where $s=z$ or $\bar{z}$. The choice will depend on the helicity of the particle and $\frac{1}{\sqrt{2}}\partial_{s}\hat{q}^{\mu}$ can be regarded as the polarization vector. The convention of the polarization vectors will be given in subsection \ref{3g}. Note that we have $X^{\mu}A_{\mu s}^{\Delta,\pm}=0$ and $\partial^{\mu}A_{\mu s}^{\Delta,\pm}=0$. The shadow conformal primary wave function is given by
\ie
\tilde{A}_{\mu s}^{\Delta,\pm}(\hat{q};X)=(-X^2)^{\Delta-1}A_{\mu s}^{\Delta,\pm}(\hat{q};X)
\fe
For amplitudes, whose legs are all on-shell, using $A_{\mu s}^{\Delta,\pm}(\hat{q};X)$ is equivalent to Mellin transformation since they are gauge equivalent:
\ie
A_{\mu s}^{\Delta,\pm}(\hat{q};X)=\frac{\Delta-1}{(\mp i)^{\Delta}\Gamma(\Delta+1)}\int d\omega\ \omega^{\Delta-1}e^{\pm i\omega\hat{q}\cdot X-\epsilon\omega}+\partial_{\mu}\alpha_{s}^{\Delta,\pm}(\hat{q};X)
\fe
However, in the soft region $\omega=0$, which forces $\Delta=1$ on the celestial side\footnote{To see this, we need to consider the integral over $\delta(\omega)$ from the momentum conservation delta function first when we are calculating a celestial amplitude \cite{Chang:2022seh}.}, the conformal wave functions need to be redefined.
\ie
A^{\log,\pm}_{\mu s}(\hat{q}_{1};X)=&\lim_{\Delta\to1}\partial_{\Delta}(A_{\mu s}^{\Delta,\pm}+\tilde{A}_{\mu s}^{2-\Delta,\pm})\\
=&\frac{2X_{\mu}}{X^2}\frac{\partial_{a}\hat{q}\cdot X}{-\hat{q}\cdot X\mp i\epsilon}+\partial_{\mu}\bigg(-\log(-X^2)\frac{\partial_{a}\hat{q}\cdot X}{-\hat{q}\cdot X\mp i\epsilon}\bigg)\\
=&V^{\log,\pm}_{\mu s}+\partial_{\mu}\alpha^{\log,\pm}_{s}
\fe
It is important to mention that $V^{\log,\pm}_{\mu s}$
and $\partial_{\mu}\alpha^{\log,\pm}_{s}$ do not satisfy the massless Klein-Gordon equation. Only the sum of these two terms can be regarded as an on-shell thing.

\subsection{BG current}
BG recursion is an off-shell recursion based on the Feynman rules. Using this recursion, one can obtain higher point BG currents from lower point ones. The definition of an $n$-pt BG current is an $(n+1)$-pt amplitude with one leg off-shell. By definition, amplitudes can be generated by BG currents in a rather simple way. Therefore, BG recursion is also a recursion for amplitudes and can be obtained from the equation of motion for an interaction system. In this subsection, we will show how to derive the BG recursion of the Yang-Mills (YM) theory as an example.

The equation of motion for YM theory takes the form (we have used the Lorenz gauge here):
\begin{equation}\label{eom}
	\square \mb{A}_{\mu}=-i[\mb{A}^{\nu},\mb{F}_{\mu\nu}]+i[\mb{A}^{\nu},\partial_{\nu}\mb{A}_{\mu}].
\end{equation}
Here, $\mb{A}_{\mu}=\text{A}_{\mu}^aT^a$ denotes the Lie algebra-valued gluon field, with $T^a$ being the Lie group generators. The corresponding Lie algebra-valued field strength tensor is given by $\mb{F}_{\mu\nu}=\partial_{\mu}\mb{A}_{\nu}-\partial_{\nu}\mb{A}_{\mu}-i[\mb{A}_{\mu},\mb{A}_{\nu}]$. Note that the Lorenz gauge is assumed in this case. To obtain the multi-particle solution of \eqref{eom}, we assume all particles are outgoing (particles with negative energies are allowed here) and apply the perturbiner expansion ansatz:
\begin{eqnarray}\label{ansatz}
    \mb{A}_{\mu}&=&\sum_P A_{P\mu}e^{ik_P\cdot x}T^{P}
\end{eqnarray}
where we use the notation $k$ to denote the momentum here, and $k_{P}=\sum_{i\in P}k_{i}$. Since the Lie algebra valued $\mb{A}_{\mu}$ is in the adjoint representation of some gauge group, its coefficients $A_{P\mu}$ must satisfy the shuffle identity\cite{Kleiss:1988ne,Lee:2015upy,Mafra:2015gia,Mizera:2018jbh}:
\begin{eqnarray}
	A_{P\shuffle Q}^{\mu}=0,\ \ P,Q\neq\varnothing
\end{eqnarray}
Substituting the ansatz \eqref{ansatz} to \eqref{eom}, we can obtain the recursion:
\ie\label{BG}
k_{P}^{2}A_{P\mu}=&\sum_{P=XY}(k_{X\mu}-k_{Y\mu})(A_{X}\cdot A_{Y})-2A_{X\mu}(k_{X}\cdot A_{Y})+2A_{Y\mu}(k_{Y}\cdot A_{X})\\
&+\sum_{P=XYZ}[2A_{Y\mu}(A_{X}\cdot A_{Z})-A_{Z\mu}(A_{X}\cdot A_{Y})-A_{X\mu}(A_{Y}\cdot A_{Z})]
\fe
with single-particle states
\ie\label{sps}
A_{i\mu}=\epsilon_{i\mu}.
\fe
The color-ordered tree-level amplitude can be obtained by the following equation
\ie
M(P,n+1)=\lim_{k_{P}\to0}k_{P}^{2}A_{n}\cdot A_{P}=\lim_{k_{P}\to0}k_{P}^{2}\epsilon_{n}\cdot A_{P}.
\fe
In this way, we can obtain any point color-ordered tree-level amplitudes by the single-particle states \eqref{sps} and the BG recursion \eqref{BG}.

\section{Celestial BG current}\label{sec3}
In this section, we will show how to obtain a celestial BG recursion and how to derive celestial amplitudes by this recursion. We only discuss the gluon case without loss of generality. In this paper, the normalization constant is not very important. Therefore we sometimes ignore these overall constants. 
\subsection{Recursion relation}
In this subsection, we will consider the Minkowski space. Without loss of generality, we define the $n$-pt celestial BG currents for $n$ incoming gluons ($\Delta_{i}\neq1$) and one outgoing off-shell gluon as follows:
\ie
\mathcal{A}_{P\mu}(\Delta_{i},p_{n+1})=\int \prod_{i=1}^{n} d\omega_{i}\omega_{i}^{\Delta_{i}-1}A_{P\mu}\delta^{(4)}(\sum_{i=1}^{n}\omega_{i}\hat{q}_{i}-p_{n+1})
\fe
where $P$ is a permutation of $\{1,2,\cdots,n\}$ and we set all on-shell legs incoming and the off-shell leg outgoing. Note that the case that all on-shell legs are outgoing and the off-shell leg is incoming has the same delta function as the definition above since the delta function is even. We also need to point out that the factor $1/k_{P}^{2}$ in the BG currents is set to $1/p_{n+1}^2$ and will not be integrated, where $p_{n+1}=m\hat{p}_{n+1}$ with a changeable $m$ is the momentum of the off-shell leg. This definition is equivalent to choosing $A_{\mu s}^{\Delta,\pm}(\hat{q};X)$ as the wave functions and setting the pure gauge terms to zero. This will change the expression of the celestial BG currents but will not change the value of the celestial amplitudes. If we want to define the general celestial BG current, we only need to consider $\delta^{(4)}(\sum_{i=1}^{n}k_i+k_{n+1})$ with $k_{i}=\epsilon_{i}\omega_{i}\hat{q}_{i}$ and $k_{n+1}=\epsilon_{n+1}m\hat{p}_{n+1}$, and then assign the particles with correct conformal primary wave functions or shadow conformal primary wave functions.

It is important to deal with the delta function. We have the following two identities:
\ie
\delta^{(4)}(\sum_{i=1}^{n}\omega_{i}\hat{q}_{i}-p_{n+1})=\int d^4Jd^4K\delta^{(4)}(q_{X}-J)\delta^{(4)}(q_{Y}-K)\delta^{(4)}(J+K-p_{n+1})
\fe
for $P=XY$ and
\ie
\delta^{(4)}(\sum_{i=1}^{n}\omega_{i}\hat{q}_{i}-p_{n+1})=\int d^4Jd^4Kd^4L\delta^{(4)}(q_{X}-J)\delta^{(4)}(q_{Y}-K)\delta^{(4)}(q_{Z}-L)\delta^{(4)}(J+K+L-p_{n+1}),
\fe
for $P=XYZ$, where $q_{X}=\sum_{i\in X}\omega_{i}\hat{q}_{i}$. From these two identities, we can obtain the recursion of the celestial BG currents:
\ie\label{cbgc}
p_{n+1}^2\mathcal{A}_{P\mu}=-&\sum_{P=XY}\int d^4Jd^4K \delta^{(4)}(J+K-p_{n+1})\times\\
&\bigg[\sum_{i\in X}\mathcal{A}_{Y}\cdot\mathcal{A}_{X}^i\hat{q}_{i\mu}-\sum_{j\in Y}\mathcal{A}_{X}\cdot\mathcal{A}_{Y}^j\hat{q}_{j\mu}+2\sum_{k\in Y}\mathcal{A}_{X}\cdot \hat{q}_{k}\mathcal{A}_{Y\mu}^k-2\sum_{l\in X}\mathcal{A}_{Y}\cdot \hat{q}_{l}\mathcal{A}^{l}_{X\mu}\bigg]\\
&+\sum_{P=XYZ}\int d^4Jd^4Kd^4L \delta^{(4)}(J+K+L-p_{n+1})\times\\
&\bigg[2\mathcal{A}_{Y\mu}(\mathcal{A}_{X}\cdot \mathcal{A}_{Z})-\mathcal{A}_{Z\mu}(\mathcal{A}_{X}\cdot \mathcal{A}_{Y})-\mathcal{A}_{X\mu}(\mathcal{A}_{Z}\cdot \mathcal{A}_{Y})\bigg]
\fe
where $\ma{A}_{X}^{i}$ denotes $\ma{A}_{X}$ with $\Delta_{i}\to\Delta_{i}+1$. The minus sign of the first term comes from $k=-\omega\hat{q}$ for incoming particles.

For BG currents, the initial condition is the single-particle state
\ie
A_{i\mu}=\epsilon_{i\mu}.
\fe
However, for celestial BG currents, things are more subtle. In fact, we have
\ie\label{c1}
\mathcal{A}_{i\mu}=\int d\omega_{i}\omega_i^{\Delta_{i}-1}\epsilon_{i\mu}\delta^{(4)}(\omega_{i}\hat{q}_{i}-p).
\fe
In this case, $p$ must be null from the momentum conservation. However, let us pretend that $p$ is still timelike first and find out another description of the delta function from the momentum conservation. Recall that $p$ can be written as 
\ie
p^{\mu}=\frac{m}{2y}(1+y^2+w\bar{w},w+\bar{w},-i(w-\bar{w}),1-y^2-w\bar{w})
\fe
with a changeable $m$. The support of the delta function is
\ie
\frac{m}{2y}=\omega_{i},y=0,w=z,\bar{w}=\bar{z}.
\fe
The Jacobian for the transformation from $(p^{\mu})\to(\frac{m}{2y},y,w,\bar{w})$ is
\ie
|J|=\frac{m^3}{y^2}
\fe
then the delta function is
\ie
\delta^{(4)}(\omega_{i}\hat{q}_{i}-p)=\frac{y^2}{m^3}\delta(\frac{m}{2y}-\omega_{i})\delta(y)\delta^{(2)}(w-z_{i}).
\fe
Substituting this to \eqref{c1}, we can obtain the initial condition for the celestial BG currents:
\ie\label{in}
\mathcal{A}_{i\mu}=\int d\omega_{i}\omega_i^{\Delta_{i}-1}\epsilon_{i\mu}\delta^{(4)}(\omega_{i}\hat{q}_{i}-p)=\frac{1}{4}\epsilon_{i\mu}\frac{m^{\Delta_{i}-4}}{(2y)^{\Delta_{i}-3}}\delta(y)\delta^{(2)}(w-z_i)
\fe
We can also obtain the celestial amplitudes from the celestial BG currents. The celestial amplitude of $n$ incoming gluons and 1 outgoing gluon can be obtained by assigning the shadow conformal primary wave function to the particle $n+1$:
\ie
\ma{M}(P,n+1)=\int\frac{d^3 \hat{p}_{n+1}}{\hat{p}_{n+1}^{0}}\tilde{A}^{\Delta_{n+1},-}_{\mu}(z_{n+1},\bar{z}_{n+1};p_{n+1})\lim_{p^{2}_{n+1}\to0}p_{n+1}^{2}\ma{A}^{\mu}_{P}.
\fe
where the shadow basis can be written as $\tilde{A}^{\Delta,-}_{\mu}(z,\bar{z};X)=\int d^4 p\tilde{A}^{\Delta,-}_{\mu}(z,\bar{z};p)e^{-ip\cdot X}$. In the Klein space, things are easier. There is no distinction between incoming and outgoing in Klein space since they are connected by a Lorentz transformation \cite{Pasterski:2017ylz}. Therefore we can always use Mellin transformation and do not need to consider the directions of the particles. In this case we have
\ie
\ma{M}(P,n+1)=\int d\omega_{n+1}\omega_{n+1}^{\Delta_{n+1}-1}\lim_{p^{2}_{n+1}\to0}p_{n+1}^{2}\epsilon_{n+1}\cdot\ma{A}_{P}.
\fe
The formalism for the celestial BG recursion looks rather simple. However, it is not easy to use this recursion due to lots of delta functions. In the next subsection, we will show an important example, and show how to use this recursion and obtain the celestial amplitudes through this example.

\subsection{Example: 3-pt gluon celestial amplitude}\label{3g}
In this subsection, we will show how to obtain the 3-pt gluon celestial amplitude only from \eqref{in} and the celestial BG recursion. This approach, based on recursion and the massless limit, is quite different from \cite{Pasterski:2017ylz}.

First of all, we need to derive the 2-pt celestial BG current $\ma{A}_{12\mu}$ (or equivalently, $p_{3}^{2}\ma{A}_{12\mu}$). Here we only show the full calculation of the term proportional to $(\epsilon_{1}\cdot\epsilon_{2})\hat{q}_{1\mu}$, which corresponds to the term $\ma{A}_{2}\cdot\ma{A}_{1}^{1}\hat{q}_{i\mu}$ in \eqref{cbgc}.

\ie
&-p_{3}^{2}\ma{A}_{12\mu}\supset(\epsilon_{1}\cdot\epsilon_{2})\hat{q}_{1\mu}\int d^4Jd^4K\frac{1}{16}\frac{2y_Jy^3}{m^2m_{J}|z_{12}|^2}\frac{m_{J}^{\Delta_{1}-3}}{(2y_{J})^{\Delta_{1}-2}}\frac{m_{K}^{\Delta_{2}-4}}{(2y_K)^{\Delta_2-3}}\delta^{(2)}(w_J-z_1)\delta^{(2)}(w_K-z_2)\\
&\times\delta(y_J)\delta(y_K)\delta(\frac{m_{J}}{2y_J}-\frac{2y_Jm^2}{4m_{J}|z_{12}|^2})\delta(y-\frac{2m\omega_1|z_{12}|^2}{m^2+4\omega_1^2|z_{12}|^2})\delta^{(2)}(w-\frac{m^2z_2+4\omega_1^2z_1z_{12}^2}{m^2+4\omega_1^2|z_{12}|^2})\\
&=(\epsilon_{1}\cdot\epsilon_{2})\hat{q}_{1\mu}\int da_Jdy_Jda_Kdy_K  \frac{1}{16}\frac{m_{J}^{3}}{y_J^2}\frac{m_{K}^{3}}{y_K^2}\frac{2y_Jy^3}{m^2m_{J}|z_{12}|^2}\frac{m_{J}^{\Delta_{1}-3}}{(2y_{J})^{\Delta_{1}-2}}\frac{m_{K}^{\Delta_{2}-4}}{(2y_K)^{\Delta_2-3}}\delta(y_J)\delta(y_K)\\
&\times\delta(\frac{m_{K}}{2y_K}-\frac{2y_Jm^2}{4m_{J}|z_{12}|^2})\delta(y-\frac{2m\omega_1|z_{12}|^2}{m^2+4\omega_1^2|z_{12}|^2})\delta^{(2)}(w-\frac{m^2z_2+4\omega_1^2z_1z_{12}^2}{m^2+4\omega_1^2|z_{12}|^2})\\
&=(\epsilon_{1}\cdot\epsilon_{2})\hat{q}_{1\mu}\int da_J  \frac{y^3}{m^2|z_{12}|^2}a_{J}^{\Delta_{1}-\Delta_{2}}(\frac{m^2}{4|z_{12}|^2})^{\Delta_{2}-1}\delta(y-\frac{2m\omega_1|z_{12}|^2}{m^2+4\omega_1^2|z_{12}|^2})\delta^{(2)}(w-\frac{m^2z_2+4\omega_1^2z_1z_{12}^2}{m^2+4\omega_1^2|z_{12}|^2})
\fe
where we have used the well-known parametrization for the delta function with respect to 2 null vectors and 1 timelike vector. We should point out that when we take $y_{J}\to0$, the variable $a_{J}=\frac{m_{J}}{2y}$ remains finite since when $y_{J}\to0$, $m_{J}\to0$.

After replacing the integral variable $a_{J}\to\omega_{1}$, the full 2-pt celestial BG current is
\ie\label{2bg}
-p_{3}^{2}\ma{A}_{12\mu}=
&\int d\omega_{1} \omega_{1}^{\Delta_{1}-1}(\frac{m^2}{4|z_{12}|^2\omega_{1}})^{\Delta_{2}-1}\frac{1}{m^2}\frac{1}{\omega_{1}|z_{12}|^{2}}y^3\\
&\times[(\omega_{1}\hat{q}_{1\mu}-\frac{m^2}{4\omega_{1}|z_{12}|^{2}}\hat{q}_{2\mu})(\epsilon_{1}\cdot\epsilon_{2})+2\epsilon_{2\mu}\frac{m^2}{4\omega_{1}|z_{12}|^{2}}\hat{q}_{2}\cdot\epsilon_{1}-2\epsilon_{1\mu}\omega_{1}\hat{q}_{1}\cdot\epsilon_{2}]\\
&\times\delta(y-\frac{2m\omega_1|z_{12}|^2}{m^2+4\omega_1^2|z_{12}|^2})\delta^{(2)}(w-\frac{m^2z_2+4\omega_1^2z_1|z_{12}|^2}{m^2+4\omega_1^2|z_{12}|^2})
\fe
This result is the same as the Mellin transformation of the 2-pt BG current, which verifies our recursion.

The next step is to obtain the 3-pt gluon celestial amplitude. For simplicity, but without loss of generality, we choose the helicity of 3 particles as $(---)$\footnote{Here the helicity is the 4D helicity for an incoming or outgoing particle. Note that the helicity $(---)$ will not give a zero since there is an outgoing particle and if we regard it as an incoming particle with minus energy we will get $+$ helicity.} and we set particle 3 to be outgoing.\footnote{It is worth mentioning that although changing the directions of particles will not change the wavefunctions, it will change the constraints of $z_{ij}$ as we will see later and finally affect the result.}. Our convention for the polarization vector of incoming particles is
\ie
\epsilon_{+}^{\mu}&=\frac{1}{\sqrt{2}}(\bar{z},1,-i,-\bar{z})\\
\epsilon_{-}^{\mu}&=\frac{1}{\sqrt{2}}(z,1,i,-z),
\fe
while for outgoing particles we need to exchange $z$ and $\bar{z}$. And we can obtain the 2-pt celestial BG current with particles 1 and 2 ``$-$" helicity:
\ie
-p_{3}^{2}\ma{A}_{12\mu}=
&\int d\omega_{1} \omega_{1}^{\Delta_{1}-1}(\frac{m^2}{4|z_{12}|^2\omega_{1}})^{\Delta_{2}-1}\frac{1}{m^2}\frac{1}{\omega_{1}|z_{12}|^{2}}y^3\\
&\times[-2\sqrt{2}\epsilon_{2\mu}\frac{m^2}{4\omega_{1}|z_{12}|^{2}}z_{12}-2\sqrt{2}\epsilon_{1\mu}\omega_{1}z_{12}]\\
&\times\delta(y-\frac{2m\omega_1|z_{12}|^2}{m^2+4\omega_1^2|z_{12}|^2})\delta^{(2)}(w-\frac{m^2z_2+4\omega_1^2z_1|z_{12}|^2}{m^2+4\omega_1^2|z_{12}|^2}).
\fe
When we take the massless limit, we must keep in mind that $\omega_{3}=\frac{m}{2y}$. We can rewrite the current so that there is no dependence on $y$. If we choose $z_{ij}$ to be non-vanishing real numbers and also independent of $\bar{z}_{ij}$, i.e. we consider the Klein space here, the support of the delta functions can be written as
\ie
\omega_{3}&=\omega_{1}\frac{z_{21}}{z_{23}}\\
\bar{z}_{12}&=\frac{m^2z_{23}}{4\omega_{1}^2z_{12}z_{31}}\\
\bar{z}_{13}&=-\frac{m^2\bar{z}_{23}}{4\omega^2_{1}|z_{12}|^2},
\fe
where we set $w=z_{3}$. Then we can rewrite the delta functions as follows:
\ie
&\delta(y-\frac{2m\omega_1|z_{12}|^2}{m^2+4\omega_1^2|z_{12}|^2})\delta^{(2)}(w-\frac{m^2z_2+4\omega_1^2z_1|z_{12}|^2}{m^2+4\omega_1^2|z_{12}|^2})\\
=&\frac{m}{2y^2}\frac{z_{12}\bar{z}_{12}}{|z_{13}||z_{23}|}\delta(\omega_{3}-\omega_{1}\frac{z_{21}}{z_{23}})\delta(\bar{z}_{12}-\frac{m^2z_{23}}{4\omega_{1}^2z_{12}z_{31}})\delta(\bar{z}_{13}+\frac{m^2\bar{z}_{23}}{4\omega^2_{1}|z_{12}|^2}),
\fe
where $|z_{ij}|$ is the absolute value of a real variable. We need to stop and make some comments about these delta functions here. There are some extra constraints on $z_{ij}$. For example, since $m$, $\omega_{1}$, $y$ are always positive, the sign of $z_{21}$, $z_{23}$, $z_{31}$ must be the same. Moreover, we have integrated a delta function $\delta(\omega_{2}-\frac{m^2}{4\omega_{1}|z_{12}|^2})$, which requires that 
\ie
\omega_{2}=\frac{m^2}{4\omega_{1}|z_{12}|^2}=\frac{z_{31}}{z_{23}}\omega_{1}.
\fe
For other cases, things are similar: we can always obtain the relations among $\omega_{i}$. Then the condition
\ie
0\leqslant\frac{\omega_{i}}{\omega_{1}+\omega_{2}+\omega_{3}}\leqslant1
\fe
in fact constrains the value of $z_{ij}$. Only when all these conditions are satisfied, the celestial amplitude does not vanish.

Let us continue. The 2-pt celestial BG current now can be written as
\ie
-p_{3}^{2}\ma{A}_{12\mu}=
&-\sqrt{2}\int d\omega_{1} \omega_{1}^{\Delta_{1}-1}(\frac{z_{31}}{z_{23}}\omega_{1})^{\Delta_{2}-1}\frac{1}{2\omega_{3}}\frac{1}{\omega_{1}|z_{12}|^{2}}[\epsilon_{2\mu}\frac{z_{31}}{z_{23}}\omega_{1}z_{12}+\epsilon_{1\mu}\omega_{1}z_{12}]\\
&\times\frac{|z_{12}|^2}{|z_{13}||z_{23}|}\delta(\omega_{3}-\omega_{1}\frac{z_{21}}{z_{23}})\delta(\bar{z}_{12}-\frac{m^2z_{23}}{4\omega_{1}^2z_{12}z_{31}})\delta(\bar{z}_{13}+\frac{m^2\bar{z}_{23}}{4\omega^2_{1}z_{12}\bar{z}_{12}}).
\fe
Now we can take the massless limit $m\to0$ (Note that $\epsilon_{+}\cdot\epsilon_{-}=1)$: 
\ie
\ma{M}_{---}(123)=&-\int d\omega_{3}\omega_{3}^{\Delta_{3}-1}\lim_{p^2_{3}\to0}p_{3}^2\epsilon_{3}\cdot\ma{A}_{12}\\
=&-\frac{\sqrt{2}}{2}\int d\omega_{1} (\omega_{1}\frac{z_{21}}{z_{23}})^{\Delta_{3}-2}\omega_{1}^{\Delta_{1}-1}(\frac{z_{31}}{z_{23}}\omega_{1})^{\Delta_{2}-1}\frac{z^2_{12}}{z_{23}}\frac{1}{|z_{13}||z_{23}|}\delta(\bar{z}_{12})\delta(\bar{z}_{13}).
\fe
Note that when we take $m\to0$, we have $\bar{z}_{23}/\bar{z}_{12}\to1$. Let $\omega_{i}=1+i\lambda_{i}$, then from the identity
\ie
\int d\omega \omega^{i\lambda-1}=2\pi\delta(\lambda),
\fe
we have
\ie
\ma{M}_{---}(123)=&-\frac{\sqrt{2}\pi\delta(\sum_{i}\lambda_{i}) \delta(\bar{z}_{12})\delta(\bar{z}_{13})}{z_{21}^{-\Delta_{3}}z_{23}^{\Delta_{2}+\Delta_{3}-2}z_{31}^{1-\Delta_{2}}|z_{13}||z_{23}|}\\
=&\frac{\sqrt{2}\pi\delta(\sum_{i}\lambda_{i}) \text{sgn}(z_{12}z_{23}z_{31})\delta(\bar{z}_{12})\delta(\bar{z}_{13})}{|z_{12}|^{-\Delta_{3}}|z_{23}|^{2-\Delta_{1}}|z_{13}|^{2-\Delta_{2}}}.
\fe
This result matches with \cite{Pasterski:2017ylz} up to some normalizations and conventions.

\subsection{Soft region}
The delta function $\delta^{(4)}(q_{1}+q_{2}-p_{3})$ has another support
\ie
\omega_{1}&=0\\
q_{2}&=p_{3}.
\fe
This corresponds to the soft region, with $\Delta_{1}$ forced to be 1 \cite{Chang:2022seh}. However, the wave function of a soft gluon is not equivalent to the Mellin transformation on plane waves. Therefore, We need to rewrite the celestial BG currents and consider the soft region carefully. In this subsection, we will consider the Minkowski space.

For the incoming soft gluon, i.e. particle 1, the 1-pt celestial BG current is
\ie
\ma{A}_{1\mu s}^{\text{soft}}=\int d^4 X A^{\log,-}_{\mu s}(\hat{q}_{1};X)e^{ip\cdot X}
\fe
while for particle 2, the 1-pt celestial BG current is still
\ie
\mathcal{A}_{2\mu s}=\int d\omega_{2}\omega_2^{\Delta_{2}-1}\epsilon_{2\mu}\delta(\omega_{2}q_{2}-p)=\frac{1}{4}\epsilon_{2\mu}\frac{m^{\Delta_{2}-4}}{(2y)^{\Delta_{2}-3}}\delta(y)\delta^{(2)}(w-z_2).
\fe
We can rewrite this current through a more convenient formalism (up to an overall constant):
\ie
\mathcal{A}_{2\mu}=\int d^4 X A_{\mu s}^{\Delta_{2},-}(\hat{q}_{2};X)e^{ip\cdot X}
\fe
Here we only show the concrete calculation of the first term of $p_{3}^2\mathcal{A}_{12\mu}$:
\ie
-p_{3}^2\mathcal{A}_{12\mu}\supset&\int d^4Jd^4K \delta^{(4)}(-J-K+p_{3})\int d^4 Y A^{\Delta_{2},-}_{s_{2}}(\hat{q}_{2};Y)e^{iK\cdot Y}\cdot\int d^4 X A^{\log,-}_{s_{1}}(\hat{q}_{1};X)e^{iJ\cdot X}q_{1\mu}\\
=&\int d^4Jd^4Kd^4L e^{i(-J-K+p_{3})\cdot L}\int d^4 Y A^{\Delta_{2},-}_{s_{2}}(\hat{q}_{2};Y)e^{iK\cdot Y}\cdot\int d^4 X A^{\log,-}_{s_{1}}(\hat{q}_{1};X)e^{iJ\cdot X}q_{1\mu}\\
=&i\int d^4L \delta^{(4)}(L-X)\delta^{(4)}(L-Y)e^{ip_{3}\cdot L}\int d^4 Y A^{\Delta_{2},-}_{s_{2}}(\hat{q}_{2};Y)\cdot\int d^4 X \partial_{\mu}A^{\log,-}_{s_{1}}(\hat{q}_{1};X)\\
=&i\int d^4L e^{ip_{3}\cdot L} A^{\Delta_{2},-}_{s_{2}}(\hat{q}_{2};L)\cdot\partial_{\mu}A^{\log,-}_{s_{1}}(\hat{q}_{1};L)
\fe

After some calculations, we can obtain the following result:
\ie
-p_{3}^{2}\ma{A}_{12\mu}=&i\int d^4 X[(\partial_{\mu}A^{\log,-}(\hat{q}_{1};X))\cdot A^{\Delta_{2},-}(\hat{q}_{2};X))-(A^{\log,-}(\hat{q}_{1};X))\cdot \partial_{\mu}A^{\Delta_{2},-}(\hat{q}_{2};X))\\
&+2(\partial^{\nu}A^{\Delta_{2},-}_{\mu})(\hat{q}_{2};X) A^{\log,-}_{\nu}(\hat{q}_{1};X)-2(\partial^{\nu}A^{\log,-}_{\mu}(\hat{q}_{1};X)) A^{\Delta_{2},-}_{\nu}(\hat{q}_{2};X)]e^{ip\cdot X},
\fe
where we have omitted the label $s_{i}$ to simplify the notation. Integrate this integral by part, we have
\ie
-p_{3}^{2}\ma{A}_{12\mu}=
&i\int d^4 X[(\partial_{\mu}V^{\log,-}(\hat{q}_{1};X))\cdot A^{\Delta_{2},-}(\hat{q}_{2};X))-(V^{\log,-}(\hat{q}_{1};X))\cdot \partial_{\mu}A^{\Delta_{2},-}(\hat{q}_{2};X))\\
&+(\partial^{\nu}A^{\Delta_{2},-}_{\mu})(\hat{q}_{2};X) V^{\log,-}_{\nu}(\hat{q}_{1};X)-ip^{\nu}A^{\Delta_{2},-}_{\mu}V^{\log,-}_{\nu}-2(\partial^{\nu}V^{\log,-}_{\mu}(\hat{q}_{1};X)) A^{\Delta_{2},-}_{\nu}(\hat{q}_{2};X)\\
&+ip_{\mu}A^{\Delta_{2},-}_{\nu}\partial^{\nu}\alpha^{\log,-}-p^2A^{\Delta_{2},-}_{\mu}\alpha^{\log,-}]e^{ip\cdot X}.
\fe
After some calculation, we find that the sum of the terms
\ie
\partial_{\mu}V^{\log,-}\cdot A^{\Delta_{2},-}-V^{\log,-}\cdot \partial_{\mu}A^{\Delta_{2},-}=&\frac{2\epsilon_{1}\cdot X}{X^2(-\hat{q}_{1}\cdot X)}A^{\Delta_{2},-}_{\mu}+\frac{2\epsilon_{1}\cdot X}{X^2(-\hat{q}_{1}\cdot X)}A^{\Delta_{2},-}_{\mu}\\
=&\frac{4\epsilon_{1}\cdot X}{X^2(-\hat{q}_{1}\cdot X)}A^{\Delta_{2},-}_{\mu}
\fe
and
\ie
-2\partial^{\nu}V^{\log,-}_{\mu} A^{\Delta_{2},-}_{\nu}=-\frac{4\epsilon_{1}\cdot X}{X^2(-\hat{q}_{1}\cdot X)}A^{\Delta_{2},-}_{\mu}-\frac{4X_{\mu}}{X^2(-\hat{q}_{1}\cdot X)}\epsilon_{1}\cdot A^{\Delta_{2},-}-\frac{4X_{\mu}\epsilon_{1}\cdot X}{X^2(-\hat{q}_{1}\cdot X)}\hat{q}_{1}\cdot A^{\Delta_{2},-}
\fe
will vanish after we consider the on-shell wave function of particle 3. If fact, all the following terms will vanish after we take the on-shell limit
\ie
ip_{\mu}A^{\Delta_{2},-}_{\nu}\partial^{\nu}\alpha^{\log,-}-p^2A^{\Delta_{2},-}_{\mu}\alpha^{\log,-}-\frac{4X_{\mu}}{X^2(-\hat{q}_{1}\cdot X)}\epsilon_{1}\cdot A^{\Delta_{2},-}-\frac{4X_{\mu}\epsilon_{1}\cdot X}{X^2(-\hat{q}_{1}\cdot X)}\hat{q}_{1}\cdot A^{\Delta_{2},-},
\fe
since if we take particle 3 on-shell, we have $p^2=0$, $X_{\mu}\cdot\tilde{A}^{\Delta_{3}}=0$ and $\partial^{\mu}\tilde{A}^{\Delta_{3}}_{\mu}=0$ for the on-shell wave function of particle 3.

If we ignore these terms, we will get
\ie
p_{3}^{2}\ma{A}_{12\mu}=
&-i\int d^4 X[(\partial^{\nu}A^{\Delta_{2},-}_{\mu}) V^{\log,-}_{\nu}-ip^{\nu}A^{\Delta_{2},-}_{\mu}V^{\log,-}_{\nu}]e^{ip\cdot X}
\fe
and the celestial amplitude is given by
\ie
\ma{M}=\lim_{p_{3}^2\to0}\int dp_{3}\ p_{3}^2\ma{A}_{12}\cdot\tilde{A}^{\Delta_{3},+}(\hat{p}_{3};p_{3})
\fe
where $\tilde{A}^{\Delta,\pm}(z,\bar{z};p)$ is defined as $\int dp\ \tilde{A}^{\Delta,\pm}(z,\bar{z};p)e^{\pm ip\cdot X}=\tilde{A}^{\Delta,\pm}(z,\bar{z};X)$.

This result matches with \cite{Chang:2022seh} up to some normalizations after we contract the celestial BG current with the on-shell wave function of particle 3. Assigning the helicity of 3 particles $(+-+)$, we write down the final result directly:
\ie
M_{1^+_{\text{soft}}2^-\to 3^+}\propto \delta(\Delta_{2}-\Delta_{3})\frac{1}{z_{23}^{\Delta_{2}-1}\bar{z}_{23}^{\Delta_{2}+1}}(\frac{1}{z_{12}}+\frac{1}{z_{31}}).
\fe

\subsection{The OPE limit of celestial BG currents}
In celestial BG currents, we can do the OPE for the legs on the celestial sphere. In this subsection, we will focus on the holomorphic collinear limit $z_{1}\to z_{2}$. Consider the 2-pt celestial BG current \eqref{2bg}, we set the helicity to be $(++)$ in order to obtain a singularity of $z_{12}$:
\ie
p_{3}^{2}\ma{A}_{12\mu}=
&2\sqrt{2}\int d\omega_{1} \omega_{1}^{\Delta_{1}-1}(\frac{m^2}{4|z_{12}|^2\omega_{1}})^{\Delta_{2}-1}\frac{1}{m^2}\frac{1}{\omega_{1}z_{12}}y^3[\epsilon_{2\mu}\frac{m^2}{4\omega_{1}|z_{12}|^{2}}+\epsilon_{1\mu}\omega_{1}]\\
&\times\delta(y-\frac{2m\omega_1|z_{12}|^2}{m^2+4\omega_1^2|z_{12}|^2})\delta^{(2)}(w-\frac{m^2z_2+4\omega_1^2z_1|z_{12}|^2}{m^2+4\omega_1^2|z_{12}|^2})
\fe
We can solve $\omega_{1}$ from the delta function $\delta(y-\frac{2m\omega_1|z_{12}|^2}{m^2+4\omega_1^2|z_{12}|^2})$:
\ie
\omega_{1}=\frac{\omega_{3}}{2}\pm\frac{1}{2}\sqrt{\omega_{3}^2-\frac{m^2}{|z_{12}|^2}}=\omega_{\pm}>0
\fe
Then we have
\ie
p_{3}^{2}\ma{A}_{12\mu}=
&2 \omega_{+}^{\Delta_{1}-1}(\omega_{-})^{\Delta_{2}-1}\frac{1}{\omega_{3}}\frac{\omega_{+}}{4|\omega_{+}-\omega_{-}|}\frac{1}{\omega_{+}z_{12}}[\epsilon_{2\mu}\omega_{-}+\epsilon_{1\mu}\omega_{+}]\\
&\times\delta^{(2)}(w-\frac{m^2z_2+4\omega_+^2z_1|z_{12}|^2}{m^2+4\omega_+^2|z_{12}|^2})+(+\leftrightarrow-)
\fe
Now we take the holomorphic collinear limit $z_{1}\to z_{2}$:
\ie
p_{3}^{2}\ma{A}_{12\mu}\sim
& \sqrt{2}\omega_{+}^{\Delta_{1}-1}\omega_{-}^{\Delta_{2}-1}\frac{1}{2|\omega_{+}-\omega_{-}|}\frac{1}{z_{12}}\epsilon_{2\mu}\delta^{(2)}(w-z_{2})+(+\leftrightarrow-)
\fe
Let $\frac{w_{-}}{w_{+}}=t$, then
\ie\label{tope}
\ma{A}_{12\mu}\sim
&-\frac{t^{\Delta_{1}-1}+t^{\Delta_{2}-1}}{\sqrt{2}|1-t|} \frac{\omega_{+}^{\Delta_{1}+\Delta_{2}-3}}{m^2}\frac{1}{z_{12}}\epsilon_{2\mu}\delta^{(2)}(w-z_{2})
\fe
where we have used $p_{3}^2=-m^2$. 

The above calculation implies that the OPE behavior of celestial BG currents is full of subtleties. However, we can still extract some useful information from it. Recall that the 1-pt celestial BG current for particle 2 is given by
\ie
\mathcal{A}_{2\mu}=\frac{1}{4}\epsilon_{2\mu}\frac{m^{\Delta_{2}-4}}{(2y)^{\Delta_{2}-3}}\delta(y)\delta^{(2)}(w-z_2)=\frac{1}{2}\epsilon_{2\mu}\frac{\omega^{\Delta_{2}-2}}{m^2}y\delta(y)\delta^{(2)}(w-z_2)
\fe
where we set $w=\frac{m}{2y}$. Note that $y\delta(y)$ always vanishes. However, in the 1-pt celestial BG current, the momentum conservation implies $\omega_{2}=\omega$. If we also impose this condition for the 2-pt celestial BG current, i.e. we impose $\omega_{+}=\omega_{3}$ for \eqref{tope}, this condition will correspond to $t=0$ and also brings a vanishing factor. This fact tells us that we will find the 1-pt celestial BG current with conformal dimension $\Delta_{1}+\Delta_{2}-1$ embedded in the 2-pt celestial BG currents after taking $z_{12}\to0$. This feature from the OPE limit is the same as \cite{Pate:2019lpp}, where they considered the OPE of two gluon operators.

\section{From tree to loop}\label{sec4}
For BG currents, there exists a technique called the ``sewing procedure" \cite{Gomez:2022dzk}. Using this technique, one can obtain 1-loop integrands from BG currents. In this section, we will generalize this technique to celestial BG currents, and summarize how to get 1-loop integrands from celestial BG currents directly. In this section, we still use $k=\epsilon\omega \hat{q}$ to denote the momentum vector in BG recursion. To make things easier, in this section we will consider the Klein space, where for all on-shell particles we can use the Mellin transformation. In order not to lose the off-shell information, we will use the Feynman gauge in this section.

\begin{figure}[H]
	\centering
	\begin{tikzpicture}[line width=1pt,scale=1.5]
		\draw[photon] (2.5,0.5)--(3.6,0);
		\draw[photon,fill=white] (4,0) circle (.5cm);
		\draw[photon] (4.5,0)--(5.6,0.5);
        \draw[photon] (2.5,-0.5)--(3.6,0);
        \draw[photon] (4.5,0)--(5.6,-0.5);
	\end{tikzpicture}
	\caption{The repeated diagram}
 \label{re}
\end{figure}
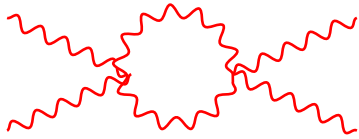

Let us start with the BG recursion \eqref{BG}. Now we replace the word $P$ with $lP$ and the word $P$ in $lP$ is still a permutation of $\{1,2,\cdots,n)$. In other words, we consider the (n+1)-pt BG current with the first on-shell leg labeled by $l$. The leg $l$ will be taken to be off-shell later so that we can ``sew" it with the original off-shell leg which is labeled by $n+1$.
\ie\label{lbg}
&k_{lP}^2A_{lP\mu}=(k_{l\mu}-k_{P\mu})(A_{l}\cdot A_{P})-A_{l\mu}(k_{l}\cdot A_{P})-A_{l\mu}(k_{lP}\cdot A_{P})+A_{P\mu}(k_{P}\cdot A_{l})+A_{P\mu}(k_{lP}\cdot A_{l})\\
&+\sum_{P=XY}[2A_{X\mu}(A_{l}\cdot A_{Y})-A_{Y\mu}(A_{l}\cdot A_{X})-A_{l\mu}(A_{X}\cdot A_{Y})]\\
&+\sum_{P=XY}(k_{lX\mu}-k_{Y\mu})(A_{lX}\cdot A_{Y})-A_{lX\mu}(k_{lX}\cdot A_{Y})-A_{lX\mu}(k_{lP}\cdot A_{Y})+A_{Y\mu}(k_{Y}\cdot A_{lX})+A_{Y\mu}(k_{lP}\cdot A_{lX})\\
&+\sum_{P=XYZ}[2A_{Y\mu}(A_{lX}\cdot A_{Z})-A_{Z\mu}(A_{lX}\cdot A_{Y})-A_{lX\mu}(A_{Z}\cdot A_{Y})].
\fe
Now we take the leg $l$ to be off-shell. This operation does nothing on the multi-point BG currents but only tells us that $k_{l}^{2}\neq0$. The first two lines of \eqref{lbg} will correspond to the tadpole terms after the sewing procedure and will vanish finally after integrating the loop momentum. Therefore, we can ignore such terms. Note that for a subset $X$ of $P$ ($X\neq P$), we cannot do the same thing for the recursion of $A_{lX}$ since this case does not appear tadpole terms. The remaining terms are
\ie
&k_{lP}^2A_{lP\mu}=\sum_{P=XY}(k_{lX\mu}-k_{Y\mu})(A_{lX}\cdot A_{Y})-A_{lX\mu}(k_{lX}\cdot A_{Y})-A_{lX\mu}(k_{lP}\cdot A_{Y})\\
&+A_{Y\mu}(k_{Y}\cdot A_{lX})+A_{Y\mu}(k_{lP}\cdot A_{lX})+\sum_{P=XYZ}[2A_{Y\mu}(A_{lX}\cdot A_{Z})-A_{Z\mu}(A_{lX}\cdot A_{Y})-A_{lX\mu}(A_{Z}\cdot A_{Y})].
\fe
Now we set $A_{lP\mu}=A_{l}^{\nu}J_{P\mu\nu}$, then
\ie
k_{lP}^2A_{l}^{\nu}J_{P\mu\nu}=&\sum_{P=XY}(k_{lX\mu}-k_{Y\mu})(A_{l}^{\nu}J_{X\mu\nu} A^{\mu}_{Y})-A_{l}^{\nu}J_{X\mu\nu}(k_{lX}\cdot A_{Y})+A_{Y\mu}(k_{Y}^{\mu} A_{l}^{\nu}J_{X\mu\nu})\\
&-A_{l}^{\nu}J_{X\mu\nu}(k_{lP}\cdot A_{Y})+A_{Y\mu}(k_{lP}^{\mu} A_{l}^{\nu}J_{X\mu\nu})\\
&+\sum_{P=XYZ}[2A_{Y\mu}(A_{l}^{\nu}J_{X\mu\nu}A^{\mu}_{Z})-A_{Z\mu}(A_{l}^{\nu}J_{X\mu\nu} A^{\mu}_{Y})-A_{l}^{\nu}J_{X\mu\nu}(A_{Z}\cdot A_{Y})].
\fe
Here $J_{P\mu\nu}$ is called the pre-integrand. After setting $k_{l}=-k_{n+1}=l$, we have
\ie
\eta^{\mu\nu}J_{P\mu\nu}=&\frac{1}{k_{l}^{2}}\sum_{P=XY}(k_{lX}^{\nu}-k_{Y}^{\nu})(J_{X\mu\nu} A^{\mu}_{Y})-J_{X\mu}^{\mu}(k_{lX}\cdot A_{Y})+A_{Y}^{\nu}(k_{Y}^{\mu} J_{X\mu\nu})-J_{X\mu}^{\mu}(k_{l}\cdot A_{Y})+A_{Y}^{\nu}(k_{l}^{\mu} J_{X\mu\nu})\\
&+\sum_{P=XYZ}[2A_{Y}^{\nu}(J_{X\mu\nu}A^{\mu}_{Z})-A_{Z}^{\nu}(J_{X\mu\nu} A^{\mu}_{Y})-J_{X\mu}^{\mu}(A_{Z}\cdot A_{Y})].
\fe
Then we have ``sewed" the legs $l$ and $n+1$. One may find that this operation is equivalent to replace $\epsilon_{l\mu}\epsilon_{n+1,\nu}$ with $\eta_{\mu\nu}$, which is just the Feynman rules. From $$\delta^{(4)}(k_{X}+k_{Y})=\int dJ^4dK^4 \delta^{(4)}(k_{X}-l-J)\delta^{(4)}(k_{Y}-K)\delta^{(4)}(J+K+l),$$ the celestial loop integrand can be written as
\ie
\ma{I}_{P}^{1-loop}=&\frac{1}{l^2}\sum_{P=[XY]}\int d^4Jd^4K \delta^{(4)}(J+K+l) \times \bigg[-2\ma{J}^{\mu}_{X\mu}l\cdot\ma{A}_{Y}+l^{\nu}\ma{J}_{X\mu\nu}\ma{A}_{Y}^{\mu}+l^{\mu}\ma{J}_{X\mu\nu}\ma{A}_{Y}^{\nu}\\
&+\sum_{i\in X}\mathcal{A}_{Y}^{\mu}\mathcal{J}_{X\mu\nu}^i\epsilon_{i}\hat{q}_{i}^{\nu}-\sum_{j\in Y}\mathcal{J}_{X\mu\nu}\mathcal{A}_{Y}^{j\mu}\epsilon_{j}\hat{q}_{j}^{\nu}+\sum_{k\in Y}\mathcal{J}_{X\mu\nu}\epsilon_{k}\hat{q}_{k\mu}\mathcal{A}_{Y}^{k\nu}-\sum_{m\in X}\mathcal{A}_{Y}\cdot \epsilon_{m}\hat{q}_{m}\mathcal{J}_{X\mu}^{m\mu}\bigg]\\
&+\frac{1}{l^2}\sum_{P=[XYZ]}\int d^4Jd^4Kd^4L \delta^{(4)}(J+K+L+l)\\
&\times\bigg[2\ma{A}_{Y}^{\nu}(\ma{J}_{X\mu\nu}A^{\mu}_{Z})-\ma{A}_{Z}^{\nu}(\ma{J}_{X\mu\nu} \ma{A}^{\mu}_{Y})-\ma{J}_{X\mu}^{\mu}(\ma{A}_{Z}\cdot \ma{A}_{Y})\bigg]
\fe
where
\ie
\ma{J}_{P\mu\nu}=\int \prod_{i\in P} d\omega_{i}\omega_{i}^{\Delta_{i}-1}J_{P\mu\nu}\delta^{(4)}(\sum_{i\in P}k_{i}-p_{n+1}-l)
\fe
As before, $\ma{A}_{Y}^{j\mu}$ means $\ma{A}_{Y}^{\mu}$ with $\Delta_{j}\to\Delta_{j}+1$. Here $[\cdots]$ means all inequivalent deconcatenations of cyclic permutations in $P$. In $P=[XY]$ and $P=1234$, for example, $(1,234)$ and $(234,1)$ are equivalent. As for $P=[XYZ]$, all deconcatenations of cyclic permutations in $P$ are inequivalent. However, in this case, different elements in $P=[XYZ]$ may give the same Feynman diagrams even though they are inequivalent since one deconcatenation corresponds to many Feynman diagrams. For example, say $P=1234$, $(12,3,4)$ and $(34,1,2)$ both give the Feynman diagram in Fig. \ref{re}. When we use this prescription to calculate 1-loop integrands, we must avoid double-counting some Feynman diagrams. 

From the derivation above, we can summarize how to obtain the celestial loop integrands from the celestial BG currents directly. 
\begin{enumerate}
    \item Consider the inverse of Mellin transformation:
\ie
f(w)=\int_{c-i\infty}^{c+i\infty}d\Delta w^{-\Delta}\tilde{f}(\Delta).
\fe
Then consider $\ma{A}_{lP}$ and do the inverse Mellin transformation on the leg $l$:
\ie
\tilde{\ma{A}}_{lP}=\int_{c-i\infty}^{c+i\infty}d\Delta_{l} \omega_{l}^{-\Delta}\ma{A}_{lP}
\fe
\item let $\tilde{\ma{A}}_{lP\mu}=\epsilon_{l}^{\nu}\ma{J}_{\mu\nu}$, then set the loop momentum $l=k_{l}=-k_{n}$.
\item consider $\ma{I}_{P}^{1-loop}=\eta^{\mu\nu}\ma{J}_{\mu\nu}$ and the modification of the deconcatenation sum $P=XY\to P=[XY]$, this is the celestial loop integrand in the Klein space.
\end{enumerate}
We have not considered the ghost loop here. However, things are similar for the ghost loop, the only subtlety is that if our external leg is a ghost, it can only be an off-shell leg. The total 1-loop integrand should be written as
\ie
I=I_{\text{gluon}}-I_{\text{ghost}}.
\fe
For more about the ghost loop from the sewing procedure, see \cite{Gomez:2022dzk}.

\section{Conclusions and outlooks}
In this paper, we have shown how to construct celestial BG currents and how to use celestial BG currents to calculate celestial amplitudes. Moreover, we can also calculate the contribution of the soft region by changing the 1-pt celestial BG current. We also explore the OPE behavior of the 2-pt celestial BG current. Finally, we have discussed the ``sewing procedure", a technique that allows us to obtain 1-loop integrands from BG currents. We have generalized this prescription to the celestial case. 

Celestial BG currents have many advantages. First, we can obtain celestial BG currents by recursion and can obtain celestial amplitudes easily from them, which means that we can calculate celestial amplitudes recursively. Moreover, compared with the BCFW recursion, BG recursion does not need to deal with the possible boundary terms. Secondly, celestial BG currents include one leg off-shell which corresponds to some delta functions involving an effective mass of the off-shell leg. This gives a new way to deal with the delta functions. Finally, since BG currents are linked closely to the Feynman rules, which is the most fundamental thing of amplitudes, celestial BG currents will be helpful in studying the structure of celestial amplitudes from the amplitude perspective.

However, there are also some problems to solve. The greatest obstacle to evaluating BG currents is the delta function coming from the momentum conservation. At least we need to find out a way to deal with the delta function of 3 massive momenta. As for the gluon case, we also need to know how to deal with the delta function of 4 massive momenta. Another interesting question is can we find an algebra for celestial BG currents just like the case in \cite{Frost:2020eoa} for BG currents, pretending that we know nothing about the flat case and just starting with the celestial side? This may deepen our understanding of the celestial holography. We leave these problems here and hope that some progress can be made.

\section*{Acknowledgement}
I would like to thank Chi-Ming Chang and Wen-Jie Ma for valuable discussions which help me a lot. YT is partly supported by National Key R\&D Program of China (NO. 2020YFA0713000). 
\appendix
\section{3D BCFW for celestial amplitudes}\label{A}
In this appendix, we want to show the 3d generalization of the BCFW recursion to deepen our understanding of the celestial recursion. We will follow some of the notations in \cite{Hu:2022bpa}. Here we only consider massless scalars and do not consider the shadow basis for simplicity, as in \cite{Lam:2017ofc}. In the 3D case, the momentum can be written as the following formalism \cite{Jiang:2022hho}:
\ie
p^{\mu}=\epsilon\omega(1+x^{2},2x,1-x^{2})
\fe
where $\epsilon$ is used to label incoming or outgoing. In 3D spinor-helicity formalism, there is no square bracket, which means we have $p^{\mu}_{ab}=\langle p|_{a}\langle p|_{b}$. We use $\lambda_{a}$ to denote $\langle p|_{a}$:
\ie
\lambda_{a}=\sqrt{-2\epsilon\omega}\left(\begin{array}{c} x\\-1  \end{array}\right)\\
\lambda^{a}=\varepsilon^{ab}\lambda_{b}=-\sqrt{-2\epsilon\omega}\left(\begin{array}{c} 1\\x  \end{array}\right)
\fe
where $\varepsilon^{ab}=\left(\begin{array}{cc} 0 & 1\\-1 & 0 \end{array}\right)$.
By chain rules and $\omega=-\frac{1}{2\epsilon}(\lambda^{1})^{2},x=\frac{\lambda^{2}}{\lambda^{1}}$, we also have
\ie
\frac{\partial}{\partial \lambda^{1}}=\frac{1}{\sqrt{-2\epsilon\omega}}(x\frac{\partial}{\partial x}-2\omega\frac{\partial}{\partial\omega})\\
\frac{\partial}{\partial \lambda^{2}}=-\frac{1}{\sqrt{-2\epsilon\omega}}\frac{\partial}{\partial x}
\fe
In the celestial case, these operators, which act directly on amplitudes, need to be translated:
\ie
\omega\frac{\partial}{\partial\omega}\rightarrow -\Delta\\
\omega\rightarrow T
\fe
where $T$ operator can add 1 to the $\Delta$ in celestial amplitudes.

In 3d BCFW, the method of the complex shift is not the same as 4D, since the constraints (momentum-conservation and on-shell condition) are so strong for 3D that the shift part must be zero. To fix this we do the nonlinear complex shift  \cite{Elvang:2013cua,Gang:2010gy}. The corresponding operator with $i,j$th particles shifted for flat amplitudes is
\ie
D_{i,j}=\exp[(\frac{z+z^{-1}}{2}-1)\lambda_{i}\frac{\partial}{\partial \lambda_{i}}-\frac{z-z^{-1}}{2i}\lambda_{j}\frac{\partial}{\partial \lambda_{i}}+(\frac{z+z^{-1}}{2}-1)\lambda_{j}\frac{\partial}{\partial \lambda_{j}}+\frac{z-z^{-1}}{2i}\lambda_{i}\frac{\partial}{\partial \lambda_{j}}]
\fe
and $D_{i,j}M=M(z)$. Here we use $M$ to denote amplitudes and $x_{ij}=x_{i}-x_{j}$. After translating to the celestial case, we have
\ie
D_{i,j}=\exp[(2-z-z^{-1})(\Delta_{i}+\Delta_{j})]\exp[\frac{z-z^{-1}}{2i}(\Omega_{i,j}(x_{ij}\frac{\partial}{\partial x_{j}}-2\Delta_{j})-\Omega_{j,i}(x_{ji}\frac{\partial}{\partial x_{i}}-2\Delta_{i}))]
\fe
where $\Omega_{i,j}=\sqrt{\frac{\epsilon_{i}}{\epsilon_{j}}}T_{i}^{1/2}T_{j}^{-1/2}$.

For 3D BCFW, we have (without loss of generality, we choose $i,j=1,n$) 
\ie
M_{n}&=\frac{1}{2\pi i}\oint_{z=1}\frac{D_{1,n}A_{n}}{z-1}\\
&=\sum_{i}[M_{L}(z_{1.i})\frac{H(z_{1,i},z_{2,i})}{P^{2}_{12\cdots i}}M(z_{1,i})+M_{L}(-z_{1.i})\frac{H(-z_{1,i},z_{2,i})}{P^{2}_{12\cdots i}}M(-z_{1,i})+(z_{1,i}\leftrightarrow z_{2,i})]
\fe
where $P^{2}_{12\cdots i}$ are propogators, $H(a,b)=\frac{a(a+1)(b^{2}-1)}{2(a^{2}-b^{2})}$, and $\pm z_{1,i},\pm z_{2,i}$ are the roots of $\hat{P}^{2}_{12\cdots i}=0$, the explicit expressions can be found in \cite{Gang:2010gy} and \cite{Elvang:2013cua}. Note that for a given $i$, $L$ includes $1,\cdots,i$ and $R$ includes $i+1,\cdots,n$.

Recall the definition of celestial amplitudes for 
massless scalars:
\ie
\mathcal{M}_{n}(\Delta_{i},x_{i})=(\prod_{i}\int_{0}^{\infty}d\omega_{i}\omega_{i}^{\Delta_{i}-1})M_{n}(\omega_{i},x_{i})\delta^{(4)}(\sum_{i}p_{i}(x_{i})).
\fe
Since $D_{i,j}$ has no effect on the delta function (it holds the momentum conservation), we have the following recursion relation for celestial amplitudes:
\ie
\mathcal{M}_{n}=&\frac{1}{2\pi i}\oint_{z=1}\frac{D_{1,n}\mathcal{M}_{n}}{z-1}\\
=&\int d^4P \sum_{i}\frac{H(z_{1,i},z_{2,i})}{P^{2}_{12\cdots i}}D_{1,n}(z_{1,i})[\ma{A}_{L}(P)\ma{A}_{R}(-P)]+\frac{H(-z_{1,i},z_{2,i})}{P^{2}_{12\cdots i}}D_{1,n}(-z_{1,i})[\ma{A}_{L}(P)\ma{A}_{R}(-P)]\\
&+(z_{1,i}\leftrightarrow z_{2,i})
\fe
where $\ma{A}(P)$ denotes the BG current (multiplied with $P^2$) with the momentum of the outgoing off-shell leg with momentum $P$. Note that $P$ can be an off-shell momentum, while $D_{1,n}P$ must be an on-shell complex momentum. This is indeed the BCFW prescription. However, this is not a recursion that constructs a celestial amplitude from lower points celestial amplitudes but lower point BG currents. To restore this, one needs to use some tricks. An example is \cite{Pasterski:2017ylz} for 4-pt MHV amplitudes. In the general case, since the operator $D_{1,n}$ has no effect on the magnitude $\omega_{P}$ of the momentum $P$ we can consider the following formalism:
\ie
&\hat{\ma{M}}(P)=D_{1,n}\ma{M}(P)=\int_{0}^{\infty} d\omega_{P}\ \omega_{P}^{\Delta_{P}-1}D_{1,n}\ma{A}(P)\\
&\rightarrow D_{1,n}\ma{A}(P)=\int_{c-i\infty}^{c+i\infty}d\Delta_{P} \omega_{P}^{-\Delta_{P}}\hat{\ma{M}}(P)
\fe
Note that we also need to figure out the shifted coordinate $x_{P}$ to obtain $\hat{\ma{M}}(P)$. After substituting this, we will obtain a recursion from lower point celestial amplitudes.

\section{Wrong conformal weights from integrating the celestial BG currents}\label{B}
An interesting trial is taking the off-shell particle to the celestial sphere. However, this prescription must be wrong because such a thing is not gauge-independent. Despite this, We can still make a calculation and see how this wrong thing manifests itself in the conformal behavior of the ``celestial amplitude".

Now we consider the 2-pt gluon BG current case. We regard the off-shell gluon as an on-shell massive spin-1 particle. In order to impose the transverse condition, we choose the Lorenz gauge. Now we define the celestial ``amplitude" for this 2-pt BG currents:
\ie\label{pca}
\tilde{\mathcal{M}}=\int \frac{dydwd\bar{w}}{y^3}(\frac{y}{y^2+|z_{3}-w_3|^2})^{\Delta_{3}+1}G_{Jb}^{(1)}p_{3}^2\mathcal{A}_{12}\cdot\epsilon_{3}^{b}
\fe
where \cite{Law:2020tsg}
\ie
G_{Jb}^{(1)}=\begin{pmatrix}
\frac{\sqrt{2}(z-w)^2}{y} & -2(z-w) & -\sqrt{2}y\\
\sqrt{2}(z-w) & \frac{|z-w|^{2}}{y} & \sqrt{2}(\bar{z}-\bar{w})\\
\sqrt{2}y & 2(\bar{z}-\bar{w}) & -\frac{\sqrt{2}(\bar{z}-\bar{w})^2}{y}.
\end{pmatrix}
\fe
We use the tilde to remind us that \eqref{pca} is not the real celestial amplitude but an experimental object. Since we have imposed the Lorenz gauge here, the polarization vectors for off-shell gluons are the same as on-shell massive spin-1 particles.
\ie
&\epsilon_{\mu}^{-}=\frac{1}{\sqrt{2}}(w,1,i,-w)\\
&\epsilon_{\mu}^{0}=-\frac{1}{2y}(1-y^2+w\bar{w},w+\bar{w},-i(w-\bar{w}),1+y^2-w\bar{w})\\
&\epsilon_{\mu}^{+}=\frac{1}{\sqrt{2}}(\bar{w},1,-i,-\bar{w})
\fe
We take the helicity of both particle 1 and particle 2 ``$-$". If we take $J=-1$, then $G_{Jb}\epsilon_{\mu}^{b}=\sqrt{2}y\epsilon^{+}_{\mu}+2(\bar{z}-\bar{w})\epsilon_{\mu}^{0}-\frac{\sqrt{2}(\bar{z}-\bar{w})^2}{y}\epsilon_{\mu}^{-}$.

After some calculation, the final result is
\ie
\tilde{\ma{M}}=&\frac{2^{-\Delta_{1}-\Delta_{2}+2}m^{\Delta_{1}+\Delta_{2}-3}\bar{z}_{23}}{|z_{12}|^{\Delta_{1}+\Delta_{2}-\Delta_{3}-1}|z_{23}|^{-\Delta_{1}+\Delta_{2}+\Delta_{3}+1}|z_{13}|^{\Delta_{1}-\Delta_{2}+\Delta_{3}+1}}B(\frac{\Delta_{1}-\Delta_{2}+\Delta_{3}+1}{2} ,\frac{-\Delta_{1}+\Delta_{2}+\Delta_{3}+1}{2} )\\
&+\frac{2^{-\Delta_{1}-\Delta_{2}+2}m^{\Delta_{1}+\Delta_{2}-3}z_{12}\bar{z}^2_{23}}{|z_{12}|^{\Delta_{1}+\Delta_{2}-\Delta_{3}+1}|z_{23}|^{-\Delta_{1}+\Delta_{2}+\Delta_{3}+3}|z_{13}|^{\Delta_{1}-\Delta_{2}+\Delta_{3}-1}}B(\frac{\Delta_{1}-\Delta_{2}+\Delta_{3}-1}{2} ,\frac{-\Delta_{1}+\Delta_{2}+\Delta_{3}+3}{2} ).
\fe
This is an unphysical result since these two terms have different conformal dimensions.

Another example is more interesting. For a 3-pt celestial ``amplitude" with 1 massless scalar (particle 1), 1 gluon (particle 2), and 1 off-shell massless scalar (particle 3), we have the result:
\ie
&\mathcal{A}_{3}=\frac{m^{\Delta_{1}+\Delta_{2}-5}|z_{23}|^{ \Delta_{12}-\Delta_{3}+1}\bar{z}_{12}}{2^{\Delta_{1}+\Delta_{2}-3/2}|z_{12}|^{\Delta_{1}+\Delta_{23}+1}|z_{13}|^{\Delta_{1}-\Delta_{23}+1}}B(\frac{\Delta_{12}+\Delta_{3}+1}{2},\frac{-\Delta_{12}+\Delta_{3}-1}{2})
\fe
It has wrong weights $(\frac{\Delta_{1}+1}{2},\frac{\Delta_{1}}{2}),(\frac{\Delta_{2}}{2},\frac{\Delta_{2}-1}{2}),(\frac{\Delta_{3}}{2},\frac{\Delta_{3}}{2})$. However, if we want this ``amplitude" to behave like a CFT correlator with correct weights, we only need to take the holomorphic collinear limit $z_{1}\to z_{2}$ so that $z_{13}=z_{23}$. In this case, from the momentum conservation, particle 3 is forced to be massless and the gauge invariant is preserved.

\bibliographystyle{JHEP}
\bibliography{cbg}

\end{document}